\documentstyle[12pt,aaspp4,flushrt]{article}

\singlespace

\font\twelvemib=cmmib10 scaled 1200 \font\tenmib=cmmib10
\font\twelvembsy=cmbsy10 scaled 1200 \font\tenmbsy=cmbsy10

\newfam\mibfam
\textfont\mibfam=\twelvemib \scriptfont\mibfam=\tenmib
\scriptscriptfont\mibfam=\tenmib
\def\mbf{\fam\mibfam}
\newfam\mbsyfam
\textfont\mbsyfam=\twelvembsy \scriptfont\mbsyfam=\tenmbsy
\scriptscriptfont\mbsyfam=\tenmbsy
\def\mbfsy{\fam\mbsyfam}

\mathchardef\alpha="710B
\mathchardef\beta="710C
\mathchardef\gamma="710D
\mathchardef\delta="710E
\mathchardef\epsilon="710F
\mathchardef\zeta="7110
\mathchardef\eta="7111
\mathchardef\theta="7112
\mathchardef\iota="7113
\mathchardef\kappa="7114
\mathchardef\lambda="7115
\mathchardef\mu="7116
\mathchardef\nu="7117
\mathchardef\xi="7118
\mathchardef\pi="7119
\mathchardef\rho="711A
\mathchardef\sigma="711B
\mathchardef\tau="711C
\mathchardef\upsilon="711D
\mathchardef\phi="711E
\mathchardef\chi="711F
\mathchardef\psi="7120
\mathchardef\omega="7121
\mathchardef\varepsilon="7122
\mathchardef\vartheta="7123
\mathchardef\varpi="7124
\mathchardef\varrho="7125
\mathchardef\varsigma="7126
\mathchardef\varphi="7127
\mathchardef\nabla="7272
\mathchardef\cdot="7201

\def\vct#1{{\mbf #1}}
\def\grad{{\mbfsy\nabla}}

\def\kms{{\rm\,km\,s^{-1}}}
\def\mpc{{\rm\,Mpc}}
\def\hmpc{{h^{-1}\rm\,Mpc}}
\def\matrixmark#1{\widehat{\cal #1}}
\def\shear{{\matrixmark S}}
\def\tshear{{\matrixmark T}}
\def\etal{{et~al.}}
\def\refrule{\makebox[3pc]{\leaders\hrule depth-2pt height 2.4pt\hfill}}

\lefthead{Kofman et al.}
\righthead{Statistics of Microlensing Magnification. I.}

\begin{document}

\hyphenation{macro-im-age mi-cro-im-age}

\title{STATISTICS OF GRAVITATIONAL MICROLENSING MAGNIFICATION.\
       I.\ TWO-DIMENSIONAL\\ LENS DISTRIBUTION}

\author{Lev Kofman}
\affil{Institute for Astronomy, University of Hawaii,
       2680 Woodlawn Drive, Honolulu, HI 96822}
\authoremail{kofman@ifa.hawaii.edu}

\author{Nick Kaiser}
\affil{CIAR Cosmology Program,
       Canadian Institute for Theoretical Astrophysics,
       University of Toronto, 60 St.~George Street, Toronto,
       Ontario M5S 1A7, Canada}
\authoremail{kaiser@cita.utoronto.ca}

\author{Man Hoi Lee}
\affil{Department of Physics, Queen's University, Kingston,
       Ontario K7L 3N6, Canada}
\authoremail{mhlee@astro.queensu.ca}

\and

\author{Arif Babul}
\affil{Department of Physics, New York University,
       4 Washington Place, New York, NY 10003}
\authoremail{babul@almuhit.physics.nyu.edu}

\bigskip

\begin{abstract}
The propagation of light of distant sources through a distribution of
clumpy matter, acting as point mass lenses, produces multiple images that
contribute to the total brightness of the observed macroimages.
In this paper we refine the theory of gravitational microlensing for a
planar distribution of point masses.
In the second accompanying paper,
we extend the analysis to a three-dimensional lens distribution.

In the two-dimensional case, we derive the probability distribution of
macroimage magnification, $P(A)$, at high magnification ($A - 1 \gg \tau^2$)
for a low optical depth ($\tau \ll 1$) lens distribution by modeling the
illumination pattern as a superposition of the patterns due to individual
``point mass plus weak shear'' lenses.
A point mass lens perturbed by weak shear $S$ produces an astroid-shaped
caustic.
We show that the magnification cross-section $\sigma(A|S)$ of the point
mass plus weak shear lens obeys a simple scaling property and provide a
useful analytic approximation for the cross-section.
By convolving this cross-section with the probability distribution of the
shear due to the neighboring point masses,
we obtain a caustic-induced feature in $P(A)$ which also exhibits a simple
scaling property.
This feature results in a 20\% enhancement in $P(A)$ at $A \approx 2/\tau$.

In the low magnification ($A - 1 \ll 1$) limit, the macroimage consists of
a single bright primary image and a large number of faint secondary images
formed close to each of the point masses.
The magnifications of the primary and the secondary images can be strongly
correlated.
Taking into account the correlations, we derive $P(A)$ for low
magnification and find that $P(A)$ has a peak of amplitude $\sim 1/\tau^2$
at $A-1 \sim \tau^2$.
The low magnification distribution matches smoothly to the high
magnification distribution in the overlapping regimes $A \ll 1/\tau$ and
$A - 1 \gg \tau^2$.

Finally, after a discussion of the correct normalization for $P(A)$,
we combine the high and low magnification results and obtain a practical
semi-analytic expression for the macroimage magnification distribution
$P(A)$.
This semi-analytic distribution is in qualitative agreement with the
results of previous numerical simulations, but the latter show stronger
caustic-induced features at moderate $A$ for $\tau$ as small as $0.1$.
We resolve this discrepancy by re-examining the criterion for low optical
depth.
A simple argument shows that the fraction of caustics of individual
lenses that merge with those of their neighbors is approximately
$1 - \exp(-8\tau)$.
For $\tau = 0.1$, the fraction is surprisingly high: $\approx 55\%$.
For the purpose of computing $P(A)$ in the manner we did, low optical depth
corresponds to $\tau \ll 1/8$.
\end{abstract}


\section{INTRODUCTION}

Gravitational lensing provides a powerful independent tool to probe the
distribution of matter in the universe and in individual astronomical objects.
There are several situations in which the lensing objects can be modeled as
an ensemble of point masses that produce multiple images (microimages) of
distant sources.  If the angular separations of the microimages are too small
to be resolved observationally, one deals with the macroimage, which is the
superposition of the microimages.
A well-known example of gravitational microlensing is the effect of individual
stars in a galaxy on the lensing properties of the galaxy (Young 1981).
If the dark matter in the galaxy is in the form of compact objects,
the optical depth for microlensing could be even higher.
Indeed, the microlensing of stars in nearby galaxies is being used to probe
for the possible existence of massive compact halo objects (MACHOs) in our
Galaxy (Paczy\'nski 1986b; Alcock \etal\ 1993; Aubourg \etal\ 1993).
Another interesting possibility is that a substantial fraction of the dark
matter in the universe is in the form of compact objects.
Then the universe as a whole has a significant optical depth to microlensing
(Press \& Gunn 1973).
Finally, microlensing can arise in a globular cluster if there are brown
dwarfs present (Paczy\'nski 1994).

The main motivation of this study is to give a systematic theory of
gravitational microlensing by a random distribution of point masses.
There are different regimes of microlensing and they often require
different types of modeling.
First, one has to distinguish the cases of low ($\tau \ll 1$),
moderate ($\tau \sim 1$), and high ($\tau \gg 1$)
optical depth, where optical depth $\tau$ is the fraction of the sky covered
(locally) by the Einstein circles of all contributing lenses:
$\tau = \pi \sum n_i x^2_{E,i}$, where $n_i$ is the projected (onto the
observer plane) surface number density of lenses with Einstein radius
$x_{E,i}$ (see \S~2).
Second, the spatial distribution of the lenses is usually either ``compact''
(e.g., stars in a distant galaxy) or ``extended'' (e.g., cosmologically
distributed compact objects).
A lens distribution is compact if the lenses are distributed in a region of
scale $R$ which is much smaller than the other distances (those between the
source and the lenses and between the observer and the lenses) in the problem;
it is extended if $R$ is comparable to the other distances in the problem.

In the case of compact lens distributions, the deflection of light rays
is generally small and essentially occurs in the vicinity of the lens
distributions.
Consequently, a compact lens distribution is usually approximated as a planar
(two-dimensional) structure corresponding to its projection
onto a lens plane that is oriented perpendicular to the line of sight
between the observer and the source.  This ``single lens plane'' approximation
allows us to pose and explore well-defined questions concerning the caustic
structure of the lens configuration, the resulting illumination pattern, the
probability distribution of macroimage magnification, the temporal
variation of the observed brightness of a source (which is the magnification
profile along a track through the illumination pattern), etc.
And while these
issues have been explored in great detail both analytically and via computer
simulations (Kaiser 1992; Mao 1992; Rauch \etal\ 1992; see also the book
by Schneider, Ehlers, \& Falco 1992 and references therein), by no means have
all the outstanding issues been resolved.

The greatest progress has been made in understanding the properties of single
plane lensing under low optical depth conditions.
One of the earliest approaches (e.g., Turner, Ostriker, \& Gott 1984)
advocated modeling the low-$\tau$ illumination pattern as a superposition
of what one would obtain for a collection of point masses acting
independently.
An isolated point mass lens produces a circularly symmetric illumination
pattern with a divergent spike at the origin.
Its differential cross-section for magnification $A$, $\sigma_0(A)$, is
proportional to the mass of the lens and scales as $A^{-3}$ for $A\gg 1$.
A simple superposition of the differential cross-sections implies that the
probability distribution of macroimage magnification $P(A)$ is
(Paczy\'nski 1986b)
\begin{equation}
P(A)\,dA = 2 \tau A^{-3} dA \qquad {\rm for~} A \gg 1,
\end{equation}
or more generally,
\begin{equation}
P(A)\,dA = 2 \tau (A^2 - 1)^{-3/2} dA \qquad {\rm for~} A - 1 \gg \tau^2.
\label{eq2}
\end{equation}

It is well known, however, that neither the $\tau$- nor the $A$-dependence is
correct in the above distributions, even in the limit of high magnification.
As shown by Nityananda \& Ostriker (1984; also Chang \& Refsdal 1984),
each lens is subject to an external shear from the macroscopic mass
distribution as well as its nearest neighbors.
An isolated point mass lens produces a degenerate point-like caustic.
External shear breaks the degeneracy and as we shall demonstrate in this
paper, the resulting caustic has the shape of an ``astroid'' and its width is
proportional to the dimensionless magnitude $S$ of the shear perturbation.
The corresponding magnification cross-section $\sigma(A|S)$ shows a strong
feature at $A \sim 1/S$.
This suggests that there ought to be a feature in $P(A)$ at $A \sim 1/\tau$
since it can readily be shown that both the typical magnitude of the
macroscopic shear and its random component are of order $\tau$.
The natural extension of the superposition approach, therefore, is to model
the illumination pattern as a superposition of the patterns due to
individual ``point mass plus weak shear'' lenses.
Briefly, the resulting probability distribution of magnification is a
convolution of the magnification cross-section $\sigma(A|S)$ for a single
lens with the probability distribution function for the shear $p(S)$.
In a subsequent section, we pursue this approach and
discuss the properties of the $P(A)$ thus acquired.

Schneider (1987a) adopted a very different approach in attempting to
determine the asymptotic behavior of $P(A)$ in the limit of high
magnification.
He argued that the high magnification events are dominated by observers
lying close to the fold caustics and hence, the illumination pattern is
dominated by two images of nearly equal brightness.
This greatly simplifies the problem.
Instead of having to identify all the microimages that make up a macroimage
and then calculate the probability distribution function for their total
magnification, one can simply use the probability distribution function
for the individual image magnification.
The latter can be calculated regardless of the optical depth.
This approach, however, has one drawback and that is the difficulty in
quantifying how large $A$ must be before $P(A)$ is well approximated by its
asymptotic form.
In the low optical depth derivation discussed above, the asymptotic form is
a good approximation for $A \gg \tau^{-1}$.
Nonetheless, it should be noted that Schneider's derivation is the only
rigorous analytic result for finite $\tau$.

Understanding the behavior of $P(A)$ at low magnification (i.e., at
$\delta A \equiv A - 1 \sim \tau^2$) is also problematic.
The distribution quoted above (eq.[\ref{eq2}]) is not properly normalized as
the integral $\int dA\,P(A)$ diverges as the lower limit of the integral
approaches $A = 1$.
One rather cavalier approach is simply to impose a cut-off at some
$A_{\rm min}$ such that the total probability is unity.
Schneider (1987b) adopted a more sophisticated approach.
He argued that a typical observer (i.e., one who is not particularly well
aligned with any single lens) will see one primary image close to the
unperturbed position of the source and a large number of faint secondary
images (one close to each lens) produced by strongly deflected rays ---
the so-called ``diffuse'' component.
Then the total macroimage magnification associated with the lensing event
is $A = A_0 + \sum_{i=1}^N A_i$,
where $A_0 = 1 + \delta A_0$ is the magnification of the primary image
and $A_i$ ($i>0$) is the magnification of the secondary images.
Assuming that $\delta A_0$ and $\sum_{i=1}^N A_i$ are statistically
independent, Schneider obtained a $P(\delta A)$ that exhibits a peak at
$\delta A \sim \tau^2$, is correctly normalized, and has the same
$A$-dependence as equation (\ref{eq2}) for $A-1 \ll 1$.
Unfortunately, the two expressions differ in {\it amplitude},
a discrepancy that arises because the magnifications of the primary and
secondary images are not independent.
In fact, we will show rigorously that the magnifications can be strongly
correlated.

In the above paragraphs, we have hinted at the various calculations and 
discussion that are to be presented in this paper.
Our overall goal is to find an expression for $P(A)$ in the low optical
depth limit that represents a synthesis of the ideas discussed above
and that is both useful as well as physically motivated. 
Our approach is a two-pronged one.
For $\delta A \gg \tau^2$, we perform the superposition of the point mass
plus weak shear lenses, allowing for the random mutual shear perturbation.
In \S\S~3.1--2 we calculate the cross-section of an individual point mass
plus weak shear lens which shows strong caustic-induced features.
In \S~3.3 we consider the random superposition of the point mass plus weak
shear lenses.
We find that the random shear tends to smear out the strong caustic-induced
features of the individual lenses and that the deviations from equation
(\ref{eq2}) around $A \sim 1/\tau$ are at the $20\%$ level.
We also find a simple scaling property for the caustic-induced ``bump'' for
low optical depth.
For low magnification, we resolve the discrepancy in Schneider's (1987b)
calculation by allowing for the correlations between the primary and
secondary images (\S~4).
This gives a formula which is valid for $A-1 \ll 1$ and which agrees with
the expression for $\delta A \gg \tau^2$ in both slope and amplitude.
An expression for $P(A)$ that includes all the relevant effects is given in
\S~5.1.
We compare this semi-analytic $P(A)$ with the numerical results of Rauch
\etal\ (1992) in \S~5.2.
While the caustic-induced bump in the semi-analytic $P(A)$ agrees
qualitatively with the bump found in the numerical simulations,
the numerical results show stronger caustic-induced features at moderate
$A$ for $\tau$ as small as $0.1$.
The additional features are due to caustic configurations more complicated
than the simple astroids of the point mass plus weak shear lenses.
To understand the contribution from the more complicated caustic
configurations at small $\tau$,
we consider the leading additional contribution from close pairs of point
masses with merged caustics.
In \S~5.3 we calculate the fraction of point masses whose caustics are not
isolated astroids as a function of $\tau$.
This gives a simple criterion for the optical depth $\tau$ below which
$P(A)$ can be constructed by the superposition of the cross sections
of individual lenses.

In the second accompanying paper (Lee \etal\ 1996; hereafter Paper II),
we conduct a similar study for the more complicated situation of a
three-dimensional distribution of point mass lenses.

\section{LENS EQUATIONS}

We begin by outlining the basic concepts of gravitational lensing,
discussing the relevant equations and defining our notation.
For the latter, we have adopted the notation of Kaiser (1992),
some of which differs slightly from those in general use.

For simplicity, we shall only consider gravitational lensing in an
Einstein-de Sitter cosmological background weakly perturbed by the
gravitational field of the lenses.
The line element for such a universe is
\begin{equation}
ds^2 = a^2(\eta) \left[(1+2\phi) d\eta^2 -
                 (1-2\phi) \left(d\chi^2 + \chi^2 d\Omega^2\right)\right] .
\end{equation}
Let the conformal time $\eta = 1$ at the present.
The scale factor is $a = a_0\eta^2$, and the Hubble parameter is
$H = H_0\eta^{-3}$ with $H_0 = 2/a_0$.
In this convention, the unit comoving length is $a_0 = 6000\hmpc$,
where $h \equiv H_0/(100 \kms\mpc^{-1})$.
The Newtonian (peculiar) potential $\phi$ induced by the density
inhomogeneities is given by Poisson's equation (in comoving coordinates):
\begin{equation}
\nabla^2\phi({\vct\chi}) = 6\eta^{-2}\Delta({\vct\chi}), \qquad
\Delta({\vct\chi}) \equiv \left(\rho({\vct\chi})-\bar\rho\right)/\bar\rho.
\end{equation}

Let us consider an isotropic point source at the origin of our coordinates
and an observer plane which is perpendicular to the $z$-axis and at a
comoving distance $\chi_{so}$ from the source (see Fig.~1).
In the absence of perturbations, the observer plane would be
uniformly illuminated and a ray which leaves the source with angle
${\vct\theta} = (\theta_1, \theta_2)$ would pierce the observer plane
at ${\vct x} = \chi_{so} {\vct \theta}$.
This defines our planar Lagrangian coordinates ${\vct x}$.
In the presence of density inhomogeneities, the light ray would suffer
deflections and the comoving displacement vector ${\vct s}({\vct x},\chi)$
at distance $\chi$ from the source is given by
\begin{equation}
{\vct s}({\vct x}, \chi) = -2 \int\limits_0^{\chi} d\chi'\;(\chi - \chi')\;
                     \grad \phi[{\vct x}' + {\vct s}({\vct x}, \chi'),\chi'],
\label{eq5}
\end{equation}
where ${\vct x}' = \chi' {\vct\theta}$.
For $\chi=\chi_{so}$, equation (\ref{eq5}) defines the mapping from
Lagrangian to Eulerian coordinates in the observer plane:
\begin{equation}
{\vct r}({\vct x}) = {\vct x} + {\vct s}({\vct x}, \chi_{so}).
\label{eq6}
\end{equation}

For most compact lens distributions of interest, the deflection of a light
ray is small and occurs essentially in the vicinity of the lens.
In effect, the deflection can be thought of as occurring at a plane that is
situated close to the location of the lens(es).
In this (single) thin-screen approximation, the lens mapping (\ref{eq6}) is a
gradient mapping with the displacement vector $\vct s = \grad\Psi$,
where $\Psi$ is an effective surface gravitational potential.
For the particular case of single plane lensing by point masses
(for simplicity, we assume that all the point masses have the {\it same}
mass $m$),
\begin{equation}
\vct s (\vct x) = -4Gm \left({\chi_{Lo}\chi_{so} \over a_L\chi_{sL}}\right)
                  \sum_k {\vct x-{\vct x}_k \over |\vct x-{\vct x}_k|^2}
                = -x^2_{E}
                  \sum_k {\vct x-{\vct x}_k \over |\vct x-{\vct x}_k|^2},
\label{eq7}
\end{equation}
where ${\vct x}_{k}$ is the position of the $k$th lens projected onto the
observer plane (i.e., ${\vct x}_k = \chi_{so} {\vct\theta}_k$ where
${\vct\theta}_k$ is the angular position of lens $k$ from the source),
$a_L$ is the scale factor at the redshift of the lens plane, and $\chi_{Lo}$,
$\chi_{so}$, and $\chi_{sL}$ are the comoving distances between the lens
plane and the observer plane, between the source and the observer plane, and
between the source and the lens plane, respectively (see Fig.~1).
In addition,
\begin{equation}
x_{E} \equiv \left(4Gm \chi_{Lo}\chi_{so}\over a_L\chi_{sL}\right)^{1/2}
\end{equation}
is the (Lagrangian) Einstein radius of a point mass on the lens plane.
For an ensemble of randomly distributed point masses, the optical depth to
microlensing is $\tau = \pi n x_{E}^2$, where
$n = (a_L \chi_{sL}/\chi_{so})^2 \Sigma/m$ is the projected (onto the
observer plane) surface number density of point masses and $\Sigma$ the
physical surface mass density of the lenses.

The generic behavior and the properties of gradient mappings such as that
defined by equation (\ref{eq6}) are relatively well-known.
In fact, the gradient mapping has been used to describe a variety of
phenomena.
The mapping ought to be familiar to the aficionados of the Zel'dovich
approximation (Zel'dovich 1970) for describing the growth of cosmological
velocities and density inhomogeneities due to gravitational instability.
The evolution of the density probability distribution function in the
Zel'dovich approximation before significant orbit crossing was studied by
Kofman \etal\ (1994).
The two-dimensional mapping has been used to describe the brightness
distribution behind a phase screen (similar to the illumination pattern
that appears at the bottom of a swimming pool) (Longuet-Higgins 1960;
Gurbatov, Malakhov, \& Saichev 1991).
In all cases, however, the results give the brightness distribution (or the
density of particles) for a single stream.
We are not aware of any multiple streaming solution for the brightness
distribution at the bottom of a swimming pool, for instance.
The situation in microlensing is quite different in that one must solve
the gradient mapping equation to locate {\it all} the microimages, and
then sum the image magnifications to obtain the macroimage magnification.
This is much more difficult.

In general, gravitational lensing maps a number of (micro)images with
Lagrangian coordinates ${\vct x}^1,\ldots,{\vct x}^n$ onto the same
Eulerian coordinate $\vct r$.
The observed flux of the image with position ${\vct x} = {\vct x}^l$ 
is magnified (or amplified) by 
$A_l = 1/|{\matrixmark D}({\vct x}^l)|$,
where
\begin{equation}
{\matrixmark D} \equiv {\partial \vct r \over \partial \vct x}
       = {\matrixmark I} + {\partial \vct s \over \partial \vct x}
\end{equation}
is the deformation tensor (or Jacobian matrix) associated with the lens
mapping and $\matrixmark I$ is the identity matrix.
If the individual microimages are not resolved observationally,
one deals with the macroimage which has a total magnification (or
amplification) $A = \sum_{l=1}^n A_l$.
In Lagrangian space ($\vct x$), the loci of points on which $|{\matrixmark D}|
= 0$
(i.e., infinite magnification) trace out curves that are called {\it critical
curves}, while the mapping of these curves to Eulerian space ($\vct r$)
defines the {\it caustics}.
In our notation, because we consider a single source and an ensemble of
observers, the caustics lie on the observer plane.
This is different from the usual practice of considering a single observer
and an ensemble of sources, which has caustics on the source plane.

For a given lens configuration, one can define a differential cross-section
$\sigma(A)$ such that $\sigma(A) dA$ is the area in the observer plane where
the total magnification is between $A$ and $A+dA$.
As we shall see, it is also useful to define a ``normalized''
differential cross-section $\varphi(A) \equiv \sigma(A)/\sigma_0(A)$,
where $\sigma_0(A)$ is the differential cross-section associated with the
simple lens configuration of a single isolated point mass (see
eq.[\ref{eq11}] below).
The function $\varphi(A)$ is a measure of the deviations of the cross-section
of a more complex lens configuration from that of the single isolated point
mass lens.
Finally, for lensing by an ensemble of randomly distributed lenses,
the generalization of the cross-section is the probability distribution of
macroimage magnification $P(A)$.

\section{CAUSTIC-INDUCED FEATURE AT HIGH MAGNIFICATION}

In \S~3.1 we solve the lens equation for a point mass lens perturbed by
weak shear.
This allows us to study analytically the caustic-induced features in the
differential cross-section (\S~3.2).
Finally, in \S~3.3, we consider the superposition of the point mass plus
weak shear lenses and derive the caustic-induced feature in $P(A)$ at high
magnification.

\subsection{``Point Mass Plus Weak Shear'' Lens: Solution of the
            Lens Equation}

The lens equation for an isolated point mass lens located at comoving
distance $\chi_{Lo}$ from the observer and at the origin of the lens plane is
\begin{equation}
\vct r = \vct x - x_E^2 {\vct x \over |\vct x|^2}.
\end{equation}
The corresponding differential cross-section is
\begin{equation}
\sigma_0(A) = {2\pi x_E^2 \over (A^2 - 1)^{3/2}} .
\label{eq11}
\end{equation}
As is well known, the caustic associated with this lens is a degenerate point
located at $\vct r = 0$, the degeneracy being due to the spherical symmetry
of the lens.

The degeneracy is lifted if the point mass is not strictly isolated and
is perturbed by shear, as in the typical lensing configuration in
single plane low optical depth microlensing.
Under low optical depth conditions, lensing (other than low magnification
events) is dominated by the point mass closest to the location where the
light ray pierces the lens plane and the contribution of the other distant
point masses can be treated as perturbations.
If we choose a coordinate system such that the dominant lens is located at
the origin of the lens plane and expand the perturbations to the first order
in $\vct x$, the perturbations correspond to a constant deflection and a
small constant shear and the lens mapping (eqs.[\ref{eq6}] and [\ref{eq7}])
simplifies to
\begin{equation}
\vct r = \vct x + {\vct s}(\vct x), \qquad
{\vct s}(\vct x) \approx {\vct d}_L(\vct x) + {\vct \alpha}_L
                         + \shear_L \vct x,
\label{eq12}
\end{equation}
where
\begin{mathletters}
\begin{equation}
{\vct d}_L (\vct x) = - x^2_{E} {\vct x\over |\vct x|^2}
\end{equation}
denotes the influence of the dominant lens,
while ${\vct \alpha}_L$ (the constant deflection) and $\shear_L$ (the shear
matrix) represent the perturbative influence of the other lenses on the plane:
\begin{eqnarray}
{\vct \alpha}_L &=& x^2_{E} \sum_k
                   {{\vct x}_k\over |{\vct x}_k|^2},
\nonumber \\ & & \label{eq13b}\\
       \shear_L &=& x^2_{E} \sum_k {1 \over |{\vct x}_k|^4}
                    \left(\begin{array}{cc}
                    x_k^2 - y_k^2&2 x_k y_k\\
                    2 x_k y_k&y_k^2 - x_k^2
                    \end{array}\right)
                 \equiv S \tshear(\phi_L).
\nonumber
\end{eqnarray}
\end{mathletters}
In the last equation, $S$ is the magnitude of the shear perturbation and
\begin{equation}
\tshear(\phi_L)=\left(\begin{array}{lr}
                \cos 2\phi_L& \sin 2\phi_L\\
                \sin 2\phi_L&-\cos 2\phi_L
                \end{array}\right) .
\end{equation}
Finally, transforming the Eulerian variable according to $\vct r \to
{\vct r}' = {\vct r} - {\vct \alpha}_L$ and rotating the coordinate axes
such that they coincide with the principal shear axes, i.e., $\phi_L = 0$,
we can cast equation (\ref{eq12}) into a more conventional form:
\begin{equation}
{\vct r}' = \vct x - x^2_{E}{\vct x\over |\vct x|^2} + S \tshear(0) \vct x,
\label{eq15}
\end{equation}
where
\begin{equation}
\tshear(0)=\left(\begin{array}{lr}
           1 &  0\\
           0 & -1
           \end{array}\right).
\end{equation}
Equation (\ref{eq15}) yields a fourth order equation for the image positions
(see Appendix A), which has four solutions inside the caustic, and two
solutions outside.

For arbitrary value of $S$, it is difficult to proceed any further,
especially with regards to determining the cross-section $\sigma(A)$
analytically.
Fortunately, the magnitude of shear is typically $S\sim \tau$ and in the
limit of low optical depth, is small (see \S~3.3).
In Figure 2 we show the caustic and the contours of constant magnification
on the observer plane for a point mass perturbed by weak shear ($S \ll 1$).
(The results shown in Figure 2 were obtained numerically for $S = 0.02$.)
For $S \ll 1$, one can observe that the entire four-stream region (i.e.,
the region inside the caustic) has high magnification.
The minimum (though high) magnification in this region occurs at
${\vct r}' = 0$ and it can be found from equations (A1)--(A3) that
\begin{equation}
A = \left(S - S^3\right)^{-1} \approx S^{-1} \gg 1
\label{eq17}
\end{equation}
at ${\vct r}' = 0$ (also Mao 1992).
Elsewhere within the caustic, $A$ is even higher and diverges to infinity at
the caustic.
This means that the light rays that delineate the four-stream region pass
very close to the critical curve in Lagrangian space.
To first order in $S$, the equation for the critical line (eq.[A4]) can be
expressed as
\begin{equation}
x^2 = x^2_{E} \left(1 + S\cos 2\vartheta\right),
\end{equation}
where we have used the polar representation $(x, \vartheta)$ for the
Lagrangian coordinate $\vct x$.
Hence, the critical line is a slightly flattened ellipse.
We can study the behavior of the rays that pass near the critical line by
considering rays with $(x/x_E)^2 = 1 + t$ and $t \ll 1$.
If we now rescale the variables: $t' = t/S$ and $\varrho' = \varrho/S =
r'/(x_E S)$, equation (A1) reduces at the lowest order in $S$ to a simple
quartic equation for $t'$:
\begin{equation}
t'^4 - (2 + \varrho'^2) t'^2 +  2 \varrho'^2 t' \cos 2\phi +
  (1 - \varrho'^2) = 0 .
\label{eq19}
\end{equation}
In general, equation (\ref{eq19}) has either two or four real solutions
(corresponding to the positions of the two images outside the caustic
or the four images inside).
Once the images are identified, their magnifications can be summed to get
the macroimage magnification factor:
\begin{equation}
A({\vct r}') = {1 \over 2S} \sum \left|t' - {(t'^2 + 1) \cos 2\phi - 2 t'
               \over t'^2 - 2 t' \cos 2\phi + 1}\right|^{-1} .
\label{eq20}
\end{equation}
Therefore, the magnification inside and around the caustic depends on the
parameter $S$ in the form $A \propto 1/S$.
{\it This means that the distribution of $A$ over the observers is a function
of the combination $SA$ only and any function of $A$ obeys a simple scaling
property.}

We can obtain a parametric expression for the caustic in Eulerian space by
solving equations (\ref{eq19}) and (\ref{eq20}):
\begin{eqnarray}
\varrho'^2(\phi) &=& {(t'^2 - 1)^2 \over t'^2 - 2 t' \cos 2\phi + 1} ,
\nonumber \\ & & \label{eq21}\\
t'(\phi) &=& \cos 2\phi - \left[\sin^2 2\phi\,(1 + \cos 2\phi)\right]^{1/3} +
             \left[\sin^2 2\phi\,(1 - \cos 2\phi)\right]^{1/3} .
\nonumber
\end{eqnarray}
After some tedious algebra, equation (\ref{eq21}) can be simplified to
\begin{equation}
\varrho'^{2/3} \left(\cos^{2/3}\phi + \sin^{2/3}\phi\right) = 2^{2/3} .
\label{eq22}
\end{equation}
Thus the caustic has the shape of an astroid, with
the four cusp catastrophes at $\phi = (n-1)\pi/2$ ($n = 1,\ldots,4$) and at
a distance $\varrho' = 2$ (or $r' = 2 x_E S$) from the center of the astroid
connected to each other by four fold catastrophes.
The caustic shown in Figure 2, which was obtained numerically for $S = 0.02$,
is described by equation (\ref{eq22}).

\subsection{Analytic Determination of the Caustic-Induced
            Features in $\sigma(A|S)$}

One consequence of the scaling relationship noted above is that the
caustic-induced features in the differential cross-section of a point mass
plus weak shear lens scale as
\begin{equation}
\sigma(A|S) = \sigma_0(A) \varphi(SA) ,
\end{equation}
where $\sigma_0(A)$ is the differential cross-section of an isolated point
mass (eq.[\ref{eq11}]) and $\varphi(SA)$ is a function describing
the scaling behavior.
In Figure 3 we show the ``normalized'' differential cross-section
$\varphi(SA)$ obtained numerically for $S = 0.02$ (solid line).
The modification of the cross-section at $SA \sim 1$ is quite pronounced
(a factor of a few higher or lower) and the deviations persist over a
decade in $SA$.
These features were first found numerically by Nityananda \& Ostriker (1984;
see also Mao 1992).
In this subsection, we investigate the properties of $\varphi(SA)$ and
find analytically the critical values of $A$ and $\varphi(SA)$.
We will also provide an analytic fit to $\varphi(SA)$.

First it can be seen that $\varphi(SA)$ tends towards unity asymptotically
at high and low magnifications.
As we have noted previously, $\sigma(A|S)\,dA$ is the area in the 
Eulerian plane between the $A$ and $A+dA$ contours.
Essentially, this is equal to the length of the iso-magnification contour
corresponding to magnification $A$ multiplied by the incremental separation
$dr'$ between the $A$ and $A+dA$ contours.
It is well known that in the vicinity of the fold caustic (i.e., in the
limit of high magnification $A \gg 1/S$), $A \propto 1/\sqrt{r_c'(\phi)-r'}$,
where $r_c'(\phi)$ is the distance to the caustic from the center of the
astroid.
On the other hand, the lengths of the iso-magnification contours are nearly
independent of magnification because both in size and in shape the contours
asymptotically approach the astroid-shaped caustic.
Consequently, $\sigma(A|S)\,dA \propto dr' \propto dA/A^3$.
Since the differential cross-section of an isolated point mass also depends
on $A$ according to $\sigma_0(A) \propto 1/A^3$ in the high magnification
limit, $\varphi(SA)$ tends to a constant value.
In fact, to first order in $S$, the magnification cross-sections of the
isolated point mass and the point mass plus weak shear cases coincide at
high magnification and $\varphi\to 1$.

At large distances from the astroid, we can solve equations (A1) and (A2)
by only considering terms to first order in $S$ and using a perturbation
series with respect to $1/r'$.
We find that the resulting iso-magnification contours are given by
$A \approx 1 + 2 (x_E/r')^4 + 2 S (x_E/r')^2 \cos 2\phi$.
In the limit of low magnification $A \ll 1/S$ (but with $A-1 \gg S^2$), the
contours are similar to those associated with an isolated point mass lens.
Consequently, the magnification cross-section converges to that of an
isolated point mass and $\varphi\to 1$.

The interesting features in $\varphi(SA)$ occur at $SA \sim 1$.
We can understand these features by studying the geometry of the
iso-magnification contours on the observer plane (see Fig.~2).
As we noted above, the iso-magnification contours at large distances from
the astroid are slightly deformed circles.
In going from low to high magnification, the quadrupole moment of the
contours increases and the iso-magnification contours are continuously
deformed.
At the first critical value $A_1$, the contour osculates the astroid and
due to symmetry, touches it at four Eulerian points with coordinates
$r'= S x_E$, $\phi = (2n-1)\pi/4$, and $n = 1,\ldots,4$.
There are four images associated with each of these points, and we can
determine their Lagrangian coordinates as well as their magnifications
from equations (\ref{eq19}) and (\ref{eq20}).
Two of these images have infinite magnification and the other two have finite
magnification; it is the latter two that are of interest.
 From equation (\ref{eq20}), we find that the total magnification associated
with the images of interest is
\begin{equation}
A_1 = 2/3\sqrt{3}S .
\end{equation}
The next contours of $A$ slightly larger than $A_1$ consist of four
symmetric arcs around and outside the four cusps and because of the loss
of the area inside the astroid the corresponding $\sigma(A)dA$ decreases
rapidly.
This explains the first break in $\varphi(SA)$ at $S A_1$.

Next, we consider the iso-magnification contours inside the caustic, i.e.,
in the four-image region.
The minimum magnification $A_2$ inside the caustic is located at $r' = 0$,
the origin of the Eulerian plane.
Using equations (\ref{eq19}) and (\ref{eq20}), we find that
\begin{equation}
A_2 = 1/S .
\label{eq25}
\end{equation}
[As we mentioned earlier, the exact solution is $A_2 = 1/S(1-S^2)$ for
$0< S <1$, which reduces to $1/S$ for $S\ll 1$; eq.(\ref{eq17})].
We can also find the magnification in the vicinity of the origin by
evaluating equations (\ref{eq19}) and (\ref{eq20}) for $r' \ll S x_E$ or
$\varrho' \ll 1$.
We find that the Eulerian position $(r',\phi)$ is associated with four
images with $t_{1,2}' = 1 \pm (r'/Sx_E) \sin\phi$ and $t_{3,4}'= -1 \pm
(r'/Sx_E) \cos\phi$ and that the total magnification is
\begin{equation}
A(r') = {1\over S} + {1\over 4S} \left({r'\over Sx_E}\right)^2 .
\end{equation}
Therefore, the differential cross-section is
\begin{equation}
\sigma(A|S) = 2 \pi r'\,{dr'\over dA}
            = 4 \pi x_E^2 S^3 \Theta(SA - 1) + \delta\sigma(A|S) ,
\label{eq27}
\end{equation}
where $\Theta$ is the step function and the term $\delta\sigma(A|S)
\sim A^{-7/2}$ (Mao 1992) takes into account the contribution to the
differential cross-section from contours with $A > A_2$ that are outside
the caustic and in the vicinity of the cusps (see Fig.~2).
The discontinuous jump at $A_2$ described by equation (\ref{eq27}) is in good
agreement with the jump of height $2$ in $\varphi(SA)$ shown in Figure 3.

Finally, we provide a simple analytic fit (with three pieces for the
different $A$ intervals) to the numerically obtained differential
cross-section of the point mass plus weak shear lens shown in Figure 3:
\begin{equation}
\begin{array}{rcll}
\varphi(SA) &=& 1 + 7.7 (SA)^{3.5} & \qquad {\rm for}~A \le A_1 ,\\
            &=& {{\textstyle 0.17} \over {\textstyle (SA - 0.33)^{1/2}}}
               +{{\textstyle 0.023} \over {\textstyle (SA - 0.33)}}
              & \qquad {\rm for}~A_1 \le A\le A_2 ,\\
            &=& 1 + {{\textstyle 0.85} \over {\textstyle SA}}
               +{{\textstyle 0.37} \over {\textstyle (SA)^5}}
              & \qquad {\rm for}~A \ge A_2 .
\end{array}
\label{eq28}
\end{equation}
This fit is shown by the dotted lines in Figure 3.

\subsection{Superposition of Cross-Sections with Distributed Shear}

Let us now consider a collection of point masses (all of the same mass $m$)
randomly distributed on a single lens plane in the low optical depth limit
($\tau \ll 1$).
The combination of individual lenses produces several important effects.
First, as we discussed previously, the lens distribution as a whole induces
shear in the neighborhoods of the individual point masses.
For a point mass perturbed by weak shear, we have found in \S\S~3.1--2 that
the shape of the iso-magnification contours has a universal form --- every
lens produces a caustic with the same astroid shape (Fig.~2) --- and that
the normalized differential cross-section $\varphi$ is a function of the
combination $SA$ only (Fig.~3).
Since the magnitude of the shear $S$ varies from lens to lens,
there is a distribution of $S$ for the cross-section
$\sigma(A|S) = \sigma_0(A)\varphi(SA)$ near each point mass.
In this subsection, we consider the superposition of the cross-sections,
with the caustic-induced features at different $A \sim 1/S$.
Other effects of the combination of individual lenses will be considered
systematically in subsequent sections.

Under low optical depth condition, the point masses are well separated,
and the caustics and the high-magnification contours on the observer plane
are relatively isolated.
Consequently, the macroimage magnification distribution $P(A)$ at high $A$
can be approximated as a superposition of the
cross-sections of the individual point mass plus weak shear lenses:
\begin{equation}
P(A) = n \int_0^\infty dS\,p(S)\,\sigma(A|S),
\label{eq29}
\end{equation}
where $p(S)$ is the probability distribution of the shear due to the other 
lenses on the plane.
For a random lens distribution with optical depth $\tau = \pi n x_E^2$,
$p(S)$ is given by (Nityananda \& Ostriker 1984; Schneider 1987b;
Lee \& Spergel 1990)
\begin{equation}
p(S) = {\tau S \over (\tau^2 + S^2)^{3/2}}.
\label{eq30}
\end{equation}
Note that the shear distribution tends to $p(S) = \tau/S^2$ for
$S \gg \tau$, which is just the shear from the nearest neighbor, and that
$p(S)$ has a prominent peak at $S=\tau/\sqrt{2}$ for $\tau \ll 1$.

Substituting equation (\ref{eq30}) into equation (\ref{eq29}), we obtain
\begin{equation}
P(A) = n\,\sigma_0(A)\,f_1(\tau A) = {2\tau \over (A^2-1)^{3/2}} f_1(\tau A),
\label{eq31}
\end{equation}
where we have introduced the function
\begin{equation}
f_1(\tau A) = \int_0^\infty dy\,{\varphi(\tau A y) y\over (1+y^2)^{3/2}}
            = 1 + \int_0^\infty dy\,{\left[\varphi(\tau A y) - 1\right] y
                                     \over (1+y^2)^{3/2}}
\label{eq32}
\end{equation}
to describe the caustic-induced feature in the macroimage magnification 
distribution.
Since $f_1(\tau A)$ depends on $\tau$ and $A$ only through the
combination $\tau A$, the caustic-induced feature in $P(A)$ also has
a simple scaling property.
The function $f_1(\tau A)$, obtained numerically with $\varphi (SA)$ as
shown in Figure 3, is plotted in Figure 4 (solid curve).
There is a mild ``bump'' located at $A \approx 2/\tau$, which is a
$20\%$ enhancement at its peak.
The bump is similar in shape to the features in $\varphi$, but it is much
weaker and smoother because of the convolution over the shear distribution.
The function $f_1(\tau A)$ provides a semi-analytic description for the
caustic-induced bump found numerically by Rauch \etal\ (1992), but only in
the low optical depth limit (see \S~5).

In Figure 4 we also show an analytic fit (dotted curve) to the
numerically evaluated $f_1(\tau A)$:
\begin{equation}
f_1(\tau A) = 1 -
    0.81 {(\tau A)^2 (1 - 3 \tau A) \over [1 + 1.5 (\tau A)^{3/2}]^{8/3}} .
\label{eq33}
\end{equation}
The maximum deviation of this fit from the numerical result is less than
$0.01$.

\section{MODIFICATION OF $P(A)$ AT LOW MAGNIFICATION}

The superposition of the cross-sections of individual lenses that results
in the caustic-induced feature $f_1(\tau A)$ is only valid for the
magnifications seen by the observers close to one of the astroid-shaped
caustics and this, as it turns out, implies the range $ A - 1 \gg \tau^2$.
Equation (\ref{eq2}) also requires modification at low magnification
(i.e., at $\delta A = A - 1 \ll 1$) as it diverges as
\begin{equation}
P(A) \approx {\tau \over \sqrt{2} (\delta A)^{3/2}}
\label{eq34}
\end{equation}
at $A = 1$ and is not normalizable.

For the modification of $P(A)$ at low magnification, we are concerned with
the observers who do not lie particularly near an astroid.
These observers see one primary image which is barely deflected from the
unperturbed position of the source and a large number of faint secondary
images, one close to each lens, which are produced by strongly deflected
rays (the diffuse component).
If we choose a coordinate system such that $\vct x = 0$ for the primary
ray,
then the lens equation near the primary ray is just equation (\ref{eq12})
without the $\vct d_L$ term,
and the magnification of the primary image (which we shall denote by $A_0$)
is obtained by summing the shear perturbations from the point masses around
the primary ray: $A_0 = (1 - S^2)^{-1} \approx 1 + S^2$, where
$S = |\vct S| = |\sum {\vct s}_k|$ and $\vct s_k$ is the shear
perturbation from the $k$th point mass in vector notation (see Appendix B).
The magnification of the faint image close to the $k$th point mass is
$A_k = x_E^4/|\vct x_k|^4 = |\vct s_k|^2$.
Hence the total excess magnification is
\begin{equation}
\delta A = A_0 + \sum_k A_k - 1 = \big|\sum_k {\vct s_k}\bigr|^2 +
                                  \sum_k \left|{\vct s_k}\right|^2 .
\label{eq35}
\end{equation}
Since the lenses have a Poisson distribution, the calculation of
$P(\delta A)$ is well defined.
Schneider (1987b) has obtained an expression for $P(\delta A)$ under the
assumption that the two terms in equation (\ref{eq35}) are statistically
independent.
While that gives a turn-over at $\delta A \sim \tau^2$,
the asymptotic behavior at $\delta A \gg \tau^2$ fails to match the
normalization of equation (\ref{eq2}).
This is not altogether surprising since for $\delta A \gg \tau^2$,
both of the sums in equation (\ref{eq35}) are dominated by the nearest lens
and the two terms are perfectly correlated.

In Appendix B, we determine $P(\delta A)$ at low magnification, taking into
account the correlations between the primary and secondary images.
This is rather complicated as $P(\delta A)$ formally reads
\begin{equation}
P(\delta A) = \int\!\!\cdots\!\!\int \prod_k d^2{\vct s}_k \, p({\vct s}_k) \,
              \delta\left[(A - 1) - \bigl|\sum_k {\vct s_k}\bigr|^2 -
                                    \sum_k \left|{\vct s_k}\right|^2
                    \right] ,
\label{eq36}
\end{equation}
where $p(\vct s_k)$ is the probability distribution of the shear from the
$k$th point mass and $\delta$ is the Dirac delta function.
Although equation (\ref{eq36}) cannot be reduced to a simple closed form,
we show in Appendix B that it can be reduced to a single integral,
which is a function of the combination $(A-1)/\tau^2$ (eq.[B14]).
Equation (B14) can be written in the form
\begin{equation}
P(A) = {\tau \over \sqrt{2}(A-1)^{3/2}}\, g_1\left[(A-1) \over \tau^2\right]
     = {2 \tau \over (A^2-1)^{3/2}}\, g_1\left[(A-1) \over \tau^2\right],
\label{eq37}
\end{equation}
where we have used the fact that $A + 1\approx 2$ for $\delta A \ll 1$.
The correction function $g_1[(A-1)/\tau^2]$ takes into account
the effect of the combination of individual lenses for small magnification.
Numerical integration of equation (B14) gives the function $g_1$ and the
distribution $P(A)$ plotted in Figure 5.
An excellent analytic fit to $g_1$ is given by
\begin{equation}
g_1(y) = {y e^{-\pi/4 y} \over \left(\pi/2^{3/2} + y \right)} .
\label{eq38}
\end{equation}
This fit has the same asymptotes (eq.[B15]) as the integral (B14),
and it does not differ from the numerical integration results by more
than $0.6\%$.

The distribution (\ref{eq37}) has the following properties. It is a normalized
distribution: $\int dA\, P(A) = 1$.
It has a sharp peak $P(A) \approx 0.16/\tau^2$ at
$\delta A \approx 0.84\tau^2$ (see Fig.~5).
For $A - 1 \gg \tau^2$, it has the asymptote
\begin{equation}
P(A) = {\tau \over \sqrt{2}(A-1)^{3/2}},
\end{equation}
which matches the low magnification ($\delta A \ll 1$) asymptote
(eq.[\ref{eq34}]) of the distribution (\ref{eq2}).
Since both distributions (\ref{eq31}) and (\ref{eq37}) give the same result
in two overlapping regimes: $A \ll 1/\tau$ and $A - 1 \gg \tau^2$,
it is easy to construct a macroimage magnification distribution that takes
into account the effects of both the caustics and the diffuse component:
\begin{equation}
P(A) = {2 \tau \over (A^2 - 1)^{3/2}}\, f_1(\tau A)\, g_1[(A-1)/\tau^2].
\label{eq40}
\end{equation}
In the appropriate limits, this distribution becomes either equation
(\ref{eq31}) or (\ref{eq37}).

\section{FINAL FORM OF $P(A)$ AND COMPARISON WITH\\
         NUMERICAL SIMULATIONS}

\subsection{Renormalization and the Final Form of $P(A)$}

In \S~3.3 we assumed that the surface density of the astroid-shaped
caustics on the observer plane is the same as the projected surface density
of point masses (see eq.[\ref{eq29}]).
There is, however, a net convergence of the light rays by the overall lens
distribution.
This increases the density of astroids by a factor $\bar A$, where
$\bar A = (1 - \tau)^{-2}$ is the average magnification on the observer plane.
Therefore, there should be an additional factor $\bar A$ in the
normalization of $P(A)$ in equation (\ref{eq31}).
In the very low optical depth limit ($\tau \ll 1$), the correction due to this 
factor $\bar A \approx 1 + 2\tau$ is small.  However, for optical depths
of order $0.1$, the correction is not negligible.

As we noted in \S~1, Schneider (1987a) derived an analytic result for
$P(A)$ in the high magnification limit, which is exact even for finite optical
depths $\tau$.
The starting point of his analysis is to consider the probability
distribution of the shear on the lens plane $p(S)$ (eq.[\ref{eq30}]).
Since the magnification for a light ray that encounters shear $S$ on the lens 
plane is $A =  |1 - S^2|^{-1}$ and each light ray corresponds to
an image in Lagrangian space,
the distribution $p(S)$ can be transformed into the probability distribution
function, $P_L(A)$, for the magnification of the individual images in
Lagrangian space.
For high magnification, $S = 1 + \delta S$ with $|\delta S| = 1/2A$, so
\begin{equation}
P_L(A) = p(S = 1) \left|d\delta S\over dA\right|
       = {\tau \over 2 (1 + \tau^2)^{3/2}}\, A^{-2} .
\end{equation}
To obtain the probability distribution function for the magnification in
Eulerian space $P(A)$,
we have to first multiply $P_L(A)$ by the factor $\bar A/A$, which takes
into account the focusing of the light rays by the lenses
(Schneider 1987b; Lee \& Spergel 1990), and then double the magnification
($A \to 2 A$), which takes into account that the high magnification events
are dominated by a pair of very bright microimages that form close to a fold
catastrophe.
The final result is
\begin{equation}
P(A) = {2 \tau {\bar A} \over (1 + \tau^2)^{3/2}} \, A^{-3}
     = {2 \tau \over (1 - \tau)^2 (1 + \tau^2)^{3/2}} \, A^{-3} ,
\label{eq42}
\end{equation}
which has the usual $A^{-3}$ form.
Although this is an exact analytic result for finite $\tau$,
it is limited in applicability to sufficiently high magnification such that
the macroimage is dominated by two microimages.
As we saw in \S~3, for low $\tau$, one actually has to go to fairly high
magnification ($A \gg 1/\tau$) for this to be a good approximation.
The normalization in equation (\ref{eq42}) differs from the normalization
$2\tau$ in equation (\ref{eq40}) by the factors $\bar A$ and
$(1 + \tau^2)^{-3/2}$.
For $\tau \la 0.1$, $\bar A \approx 1 + 2\tau$ and $(1 + \tau^2)^{-3/2}
\approx 1 - 3\tau^2/2$, and the correction to the normalization $2\tau$
is mainly due to the factor $\bar A$.
Note that we have already derived the factor $\bar A$ with the argument given
in the previous paragraph.

We can now collect together the three effects of the combination of
individual lenses discussed above --- the caustic-induced feature from
equation (\ref{eq31}), the low magnification modification due to the diffuse
component (eq.[\ref{eq37}]), and the renormalization due to the net
convergence of light rays and multiple streaming (eq.[\ref{eq42}]) --- and
find that for $\tau \ll 1$, the final form of the macroimage magnification
distribution due to a two-dimensional distribution of point masses is
\begin{equation}
P(A) = {2 \tau \over (1 - \tau)^2 (1 + \tau^2)^{3/2}} \,
       (A^2 - 1)^{-3/2} \, f_1(\tau A) \, g_1\left[(A-1)/\tau^2\right] .
\label{eq43}
\end{equation}

\subsection{Comparison with Numerical Simulations}

Several authors (e.g., Paczy\'nski 1986a; Rauch \etal\ 1992; Wambsganss 1992)
have used numerical simulations to calculate the macroimage magnification 
distribution, $P(A)$, produced by a two-dimensional distribution of point
masses.
In particular, distributions with high resolution in $A$ have been obtained
from Monte Carlo simulations by Rauch \etal\ for $\tau = 0.1$, $0.2$, and
$0.3$.
We now compare the semi-analytic $P(A)$ derived in this paper, equation
(\ref{eq43}), to these numerical results.
As noted, our derivation assumes low optical depth (i.e., $\tau \ll 1$).
It is not immediately obvious whether optical depths in the range
$0.1\leq \tau \leq 0.3$ are sufficiently small and whether we should expect
a good agreement between the semi-analytic and numerical results.
In fact, as we shall see, $\tau = 0.1$ is not sufficiently small. There
are differences between the semi-analytic and numerical results that are
indicative of finite optical depth effects.
Ideally, comparison should also be made for smaller $\tau$,
but numerical $P(A)$ with the required accuracy is not available
[the computational requirement increases rapidly with decreasing $\tau$
because the caustic-induced feature shifts to higher $A$ ($\propto 1/\tau$)
while the amplitude of $P(A)$ at large $A$ ($\propto \tau$) decreases].

In Figure 6 we show the high resolution $P(A)$ obtained by Rauch \etal\ for
$\tau = 0.1$ and $0.2$ (histograms).
The data are those shown in Figure 3 of Rauch \etal\ (1992),
but they are plotted in the form $(A^2 - 1)^{3/2} P(A)$ and the bin sizes
at large $A$ are slightly different.
For comparison, the solid lines show the full semi-analytic $P(A)$
(eq.[\ref{eq43}]), and the dotted lines show the semi-analytic $P(A)$ with
the caustic-induced feature only (i.e., eq.[\ref{eq43}] without the function
$g_1$).
Note that in a plot of $(A^2 - 1)^{3/2} P(A)$, the semi-analytic
distributions without the low magnification modification $g_1$ or the
caustic-induced feature $f_1$ are simply horizontal lines of amplitude
$2 \tau (1 - \tau)^{-2} (1 + \tau^2)^{-3/2}$ (dashed lines).

At $\log A < 0.1$ (or $\delta A < 0.3$), the semi-analytic and numerical
results for $\tau = 0.1$ are in good agreement, but the results for
$\tau = 0.2$ are slightly different in shape and amplitude.
Since this range of $A$ is dominated by the low magnification modification
$g_1$, we conclude that the function $g_1$ derived in \S~4 is valid for
$\tau < 0.2$.
(Recall that the peak in $P(A)$ produced by $g_1$ is located at
$\delta A \approx 0.84\tau^2$, which is much less than $0.3$ for both cases.)
The semi-analytic caustic-induced feature also provides a reasonably good
fit to the numerical results at $\log A > 1$ and $1.2$ for $\tau = 0.1$ and
$0.2$, respectively.
At intermediate $A$, there are significant differences between the numerical
and semi-analytic results.
The numerical distribution is lower than the semi-analytic distribution in
the neighborhood of the minimum at $A \approx 2.5$, and it is higher than
the semi-analytic distribution at smaller and larger $A$.
This pattern of deviations is similar to the cross-section shown in Figure 3
and suggests that the function $f_1$ derived in \S~3 underestimates the
strength of the caustic-induced feature at intermediate $A$.
It is, however, important to note that the differences between the
semi-analytic and numerical results decrease in strength with decreasing
$\tau$ and should be relatively small for $\tau \la 0.05$.
In the next subsection we shall analyze the cause of the differences at
intermediate $A$.

\subsection{Re-examining the Criterion for Low Optical Depth}

In our analysis of the caustic-induced feature in $P(A)$ for low optical
depth (\S~3), we assume that all the point masses are well separated.
Then the deflection near each point mass is due to that point mass and the
shear perturbation from the other lenses, and an astroid-shaped caustic is
associated with each point mass.
However, since the lenses are randomly distributed on the lens plane,
some of the lenses have close neighbor(s), separated by less than a few
Einstein radius $x_E$.
In these cases, the caustics are not isolated astroids but more complicated
structures produced by the collective effect of two (or more) point masses.
If the surface density (or optical depth) of the lens distribution is low
enough that the fraction of lenses with close neighbor(s) is
small, the contribution to $P(A)$ by these configurations is negligible,
and the analysis in \S~3 is valid.
For larger $\tau$, however, we have to take into account the more complex
configurations and can do so by evaluating the macroimage magnification
distribution as a series:
$P(A) = P_1(A) + P_2(A) + \cdots$, where $P_1(A)$ is the contribution by
the point masses perturbed by shear, $P_2(A)$ is the contribution by
close pairs of point masses, etc.
The first term $P_1(A)$ is the distribution derived in \S~3 (but with a
maximum cutoff in the convolution over shear).
The second term $P_2(A)$ can be evaluated (approximately) as a convolution
of the cross-section $\sigma(A|d)$ for two point masses separated by
distance $d$ with the probability distribution for $d$
if we consider a point mass and its nearest neighbor as a two-point-mass
lens and neglect the perturbation from the other lenses.

To understand the contribution to $P(A)$ from close pairs of point masses,
we must first look at some of the properties of lensing by two equal point
masses on a single lens plane (see \S~3 of Paper II and Schneider \& Weiss
1986 for details).
For consistency with the notation in Paper II, we shall express the
Lagrangian separation $d$ between the lenses in units of $\sqrt{2}\, x_E$.
In the limit $d \gg 2$, the region near each of the point masses is
perturbed by the weak shear ($S = 1/2 d^2$) from the other point mass,
and there are two astroid-shaped caustics.
The caustics move towards each other (and become asymmetric) as $d$ decreases,
and they touch when $d = 2$.
As $d$ decreases below $2$, the number of caustics changes from two to one
to finally three (see Fig.~3 of Paper II).
In Paper II, this sequence of caustic topologies is denoted as topology
types $A'$ (for $d > 2$), $B'$ (for $1/\sqrt{2} < d < 2$), and $C'$
(for $d < 1/\sqrt{2}$).
In all cases, the normalized differential cross-sections $\varphi(A)$ are
qualitatively similar to that for the point mass plus weak shear lens,
but the caustic-induced features can be significantly stronger (compare
Fig.~4 of Paper II to Fig.~3 of this paper).
As in the point mass plus weak shear case, there is a discontinuous jump in
$\varphi(A)$ at the minimum magnification $A_{\rm min}$ inside the
caustic(s).
In the limit $d \gg 2$, the minimum magnification inside the two
astroid-shaped caustics is $A_{\rm min} (d) \approx A_2 (S=1/2 d^2) = 2 d^2$
(eq.[\ref{eq25}]).
However, as Witt \& Mao (1995) have shown, $A_{\rm min}$ is a non-monotonic
function of $d$ and has a global minimum of $3$ when $d = \sqrt{2}$.
If we now consider the convolution of these cross-sections with the
probability distribution for $d$,
it is clear that the resulting distribution $P_2(A)$ should show an
enhancement in the caustic-induced feature near $A = 3$.

In their analysis of the caustic-induced feature in $P(A)$,
Rauch \etal\ (1992) compared the results from the full scale Monte Carlo
simulations to the results from a simpler two-point-mass model (see their
Fig.~6 for the $\tau = 0.2$ case).
Their two-point-mass calculation is in fact an approximate numerical
evaluation of $P_1(A) + P_2(A)$.
They found that the two-point-mass model is able to reproduce partly the
caustic-induced feature at intermediate $A$.
In particular, the distribution $P(A)$ from the two-point-mass model shows
a dip at $A \approx 3$.
Rauch \etal\ suggested that the slightly lower value ($A \approx 2.5$) of
the dip location found in the full Monte Carlo simulations is due to
non-negligible contributions from configurations of three or more point
masses.

Is there a simple explanation for the relatively strong contribution to
$P(A)$ by the collective effect of two (or more) point masses for $\tau$ as
small as $0.1$?
For a Poisson distribution of point masses, the probability that the
nearest neighbor to a point mass is at a distance less than $d$ (again in
units of $\sqrt{2}\, x_E$) is simply $1 - \exp(-2 \pi n x_E^2 d^2)$.
If we ignore the deflection due to the other lenses,
the point mass and its nearest neighbor is a two-point-mass lens.
As we mentioned earlier, a two-point-mass lens produces two astroid-shaped
caustics only if $d > d_{AB}$, where $d_{AB} = 2$ is the separation at
which the caustic topology changes from type $A'$ to $B'$.
Therefore, a simple estimate for the fraction of point masses whose caustics
are not isolated astroids is
\begin{equation}
P_{\rm na}(\tau) = 1 - \exp(-2 \pi n x_E^2 d_{AB}^2) = 1 - \exp(-8\tau) .
\end{equation}
For the contribution to $P(A)$ from $P_2(A)$ (and higher order terms) to be
negligible, $P_{\rm na}(\tau) \ll 1$ or $\tau \ll 1/8$.
It is immediately clear that $\tau = 0.1$ is not sufficiently small for
$P_2(A)$ to be negligible.
For $\tau = 0.1$, we estimate that about half of the point masses produce
caustic structures that are more complex than the astroid shape:
$P_{\rm na} = 0.55$.
This is consistent with the illumination pattern shown in Figure 1 of
Rauch \etal\ (1992).
In \S~5.2, we concluded from the comparison that the contribution from
$P_2(A)$ should be reasonably small for $\tau \la 0.05$;
this corresponds to $P_{\rm na} \la 0.33$.

\section{SUMMARY}

In this paper we have attempted to build upon the various approaches
developed in previous studies in order to develop a systematic theory of
gravitational microlensing by a planar distribution of point masses in the
low optical depth limit.
In particular, we have derived a practical semi-analytic expression for the
probability distribution of macroimage magnification, $P(A)$.

At high magnification ($\delta A \gg \tau^2$), we model the illumination
pattern as a superposition of the patterns due to individual ``point mass
plus weak shear'' lenses.
The shear perturbation near each point mass is induced by the neighboring point
masses.
It breaks the degeneracy of the caustic of an isolated point mass lens
and produces an astroid-shaped caustic.
The convolution of the magnification cross-section of the point mass plus
weak shear lens with the probability distribution of shear yields
$P(A) = 2\tau (A^2-1)^{-3/2} f_1(\tau A)$,
where the function $f_1(\tau A)$ (eq.[\ref{eq32}]; Fig.~4) describes the
caustic-induced feature in the macroimage magnification distribution.
In effect, $f_1(\tau A)$ introduces a mild ``bump,'' a $20\%$ enhancement,
at $A \approx 2/\tau$.
Since $f_1(\tau A)$ depends on $\tau$ and $A$ only through the combination
$\tau A$, the caustic-induced feature in $P(A)$ exhibits a simple scaling
property.
To facilitate future computations, we have provided a useful analytic fit to
$f_1(\tau A)$ (eq.[\ref{eq33}]).

We should point out that the results derived in \S\S~3.1--2 for the
point mass plus weak shear lens may also have applications in the analysis
of gravitational microlensing by physical binary systems.
Microlensing searches towards the Galactic bulge and the Large Magellanic
Cloud have already discovered microlensing events by close binaries with
a single merged caustic (Udalski \etal\ 1994; Alard, Mao, \& Guibert 1995).
Since wide binaries are more common than close ones,
there should be a significant number of events due to wide binaries
(Di Stefano \& Mao 1996).
If the components of the binary are sufficiently far apart,
the region near each point mass is perturbed by the weak shear from the
other point mass.
In these cases, there are two separate astroid-shaped caustics, and the
results derived in \S\S~3.1--2 are applicable.

At low magnification ($\delta A \ll 1$), the macroimage consists of a
bright primary image barely deflected from the unperturbed position of the
source and a large number of faint secondary images (the diffuse component)
formed close to each of the lenses.
The magnifications of the primary and the secondary images can be strongly
correlated.
Taking into account the correlations, we find that
$P(A) = 2 \tau (A^2 - 1)^{-3/2} g_1[(A-1)/\tau^2]$,
where $g_1$ (eqs.[\ref{eq37}] and [B14]; Fig.~5) represents the low
magnification correction.
An excellent analytic fit to $g_1$ is given by equation (\ref{eq38}).
The function $g_1$ prevents $P(A)$ from diverging at $A = 1$, and it
introduces a sharp peak of amplitude $P(A) \approx 0.16/\tau^2$ at
$A - 1\approx 0.84\tau^2$.

We also discussed the renormalization of $P(A)$ due to the net convergence
of light rays and multiple streaming.
Finally, collecting together the above results, we find that in the low
optical depth limit ($\tau \ll 1$),
\begin{equation}
P(A) = {2 \tau \over (1 - \tau)^2 (1 + \tau^2)^{3/2}} \,
       (A^2 - 1)^{-3/2} \, f_1(\tau A) \, g_1\left[(A-1)/\tau^2\right] .
\eqnum{\ref{eq43}}
\end{equation}
In order to determine the realm of validity of the above semi-analytic
expression,
we compared it against $P(A)$ obtained from Monte Carlo simulations by Rauch 
\etal\ (1992) for $\tau = 0.1$ and $0.2$.
At low magnifications ($\log A < 0.1$), we find that the $\tau = 0.1$
semi-analytic and numerical results are in good agreement with each other.
At greater optical depths, differences arise both in shape and amplitude.
We, therefore, conclude that the low magnification modification $g_1$ is
valid for $\tau < 0.2$.
The numerical and the semi-analytic results are also in good agreement in
the high magnification regime ($\log A > 1$ and $1.2$ for $\tau = 0.1$ and
$0.2$, respectively).  At intermediate $A$, however, the semi-analytic
expression does not match the numerical result even for optical depth 
$\tau = 0.1$.
The deviations arise because the function $f_1$ derived in \S~3
underestimates the strength of the caustic-induced feature at intermediate
$A$.
The deviations tend to diminish with decreasing optical depth,
and we expect them to be relatively small for $\tau \la 0.05$.

In order to understand the discrepancy between the numerical and semi-analytic
results for $\tau$ as small as $0.1$,
we re-examined our derivation of the function $f_1$ describing the
caustic-induced feature in $P(A)$.
In our derivation, we assume that a unique astroid-shaped caustic is
associated with each point mass.
In a random distribution of point masses, there may arise groups of two
(or more) lenses that lie sufficiently close to each other and give rise
to caustic configurations more complicated than isolated astroids.
If the surface density of point masses is small, such lens configurations
will be rare.
If, however, the surface density is large, they will be more common and
their contributions cannot be neglected.
For a distribution of optical depth $\tau$, we estimate that the fraction
of point masses whose caustics are not simple astroids is
$P_{\rm na}(\tau) = 1 - \exp(-8\tau)$.
For $\tau = 0.1$, about half of the point masses fall into this category.
For the purpose of computing the macroimage magnification distribution
$P(A)$ in the manner we did, a simple criterion for low optical depth is
$P_{\rm na}\ll 1$ or $\tau \ll 1/8$.

In the accompanying Paper II, we extend our analysis to the more complicated
situation of a three-dimensional distribution of point mass lenses.

\acknowledgments
We thank K.~Rauch for providing us with the results of Rauch \etal\ (1992).
We also thank S.~Mao and K.~Rauch for helpful discussions.
A.B.\ and L.K.\ acknowledge the hospitality shown to them at CITA during
their visits in 1995.
This work was supported in part by NSERC (Canada), the CIAR cosmology
program, CITA, the Institute for Astronomy (L.K.), the Dudley Observatory
(A.B.), and a CITA National Fellowship (M.H.L.).

\clearpage
\appendix

\section{LENS EQUATIONS FOR POINT MASS PLUS ARBITRARY SHEAR}

Expressing $\vct r'$ and $\vct x$ in polar coordinates $(r',\phi)$ and
$(x,\vartheta)$ respectively,
equation (\ref{eq15}) can be written as a fourth order equation for the
Lagrangian coordinate $\vct x$ in terms of the Eulerian coordinate $\vct r'$:
\begin{equation}
(\zeta - 1)^2 \left[\zeta^2 - (2 + \varrho^2) \zeta + 1\right] =
  -2 S \varrho^2 \cos 2\phi\,\zeta^2 (\zeta - 1) + S^2 \varrho^2 \zeta^3 +
   2 S^2 \zeta^2 (\zeta - 1)^2 - S^4 \zeta^4 ,
\end{equation}
\begin{equation}
\tan \vartheta =
  \left(1- \zeta - S \zeta \over 1 - \zeta +S \zeta\right) \tan\phi ,
\end{equation}
where we have defined $\zeta \equiv (x/x_E)^2$ and $\varrho \equiv r'/x_E$.
In this notation, the Jacobian of the mapping is
\begin{equation}
|\matrixmark D|
  = \left|\partial \vct r' \over \partial \vct x\right|
  = 1 - {1 \over \zeta^2} - {2 S \cos 2\vartheta \over \zeta} - S^2 .
\end{equation}
The critical line (i.e., the loci of Lagrangian points where the Jacobian
$|\matrixmark D|$ vanishes) is given by
\begin{equation}
\zeta = {S \over 1 - S^2} \left(\cos 2\vartheta +
        \sqrt{\cos^2 2\vartheta + {1 - S^2 \over S^2}}\right) ,
\end{equation}
which corresponds to the so-called ``Cassini oval.''
The associated caustic is non-trivial and it is not a point singularity.
Based on equations (A1) and (A2), we expect four images in the interior of
the caustic and two outside.

\section{LOW MAGNIFICATION LIMIT OF $P(A)$}

Let us consider a random distribution of $N = \pi n R^2$ point mass lenses
within a circle of Lagrangian radius $R$,
where $n$ is the projected surface number density of lenses.
The total shear at the origin due to the $N$ point masses is given in matrix
form by equation (\ref{eq13b}).
We shall find it convenient to write the two distinct components of the
shear matrix $\shear_L$ as a vector: $\vct S = \sum_{k=1}^N \vct s_k$, where
$\vct s_k = (x_E/|\vct x_k|)^2\, (\cos 2\phi_k, \sin 2\phi_k)$ is the shear
perturbation from the $k$th point mass.

To determine the low magnification limit of $P(A)$,
we need to evaluate the probability distribution function for the following
combination of shears (eq.[\ref{eq35}]):
\begin{equation}
\delta A = A - 1 = \big|\sum_{k=1}^N {\vct s_k}\bigr|^2 +
                   \sum_{k=1}^N \left|{\vct s_k}\right|^2 .
\end{equation}
The first term in equation (B1) is just $S^2 = |\vct S|^2$, and
the probability distribution of this term alone is found easily from the
distribution (\ref{eq30}).
The second term in equation (B1) is the so-called diffuse component, and
its probability distribution has previously been considered by Schneider
(1987b).
The evaluation of the probability distribution of the sum of these two terms,
which is $\delta A$, is more complicated.

Let us start with the formal expression for the probability distribution of
$\delta A$ as a function of the $N$ statistically independent shears
(eq.[\ref{eq36}]):
\begin{equation}
P(\delta A) = \int\!\!\cdots\!\!\int \prod_k d^2{\vct s}_k \, p({\vct s}_k) \,
              \delta\left[(A - 1) - \bigl|\sum_{k=1}^N {\vct s_k}\bigr|^2 -
                                    \sum_{k=1}^N \left|{\vct s_k}\right|^2
                    \right] ,
\end{equation}
where $p(\vct s_k)$ is the probability distribution of the shear from the
$k$th point mass and $\delta$ is the Dirac delta function.
Equation (B2) can be simplified if we express each term in the integrand
as a Fourier transform.
For $p(\vct s_k)$, we have
\begin{equation}
p(\vct s_k) = {1\over (2\pi)^2} \int d^2{\vct t}_k \, q({\vct t}_k) \,
              e^{-i{\vct t}_k\cdot{\vct s}_k} ,
\end{equation}
where the characteristic function $q({\vct t}_k)$ is
(Nityananda \& Ostriker 1984)
\begin{equation}
q({\vct t}_k) = 1 - {\tau \over N}t_k ,
\end{equation}
and $\tau = \pi n x_E^2$ is the optical depth.
Because of the cross terms in $|\vct S|^2$, it is inconvenient to use the
direct Fourier transform of the delta function in equation (B2).
Instead, one can introduce an auxiliary integration to express the delta
function in the form
\begin{eqnarray}
\lefteqn{
\delta\Biggl[(A - 1) - \bigl|\sum_{k=1}^N {\vct s_k}\bigr|^2
                     - \sum_{k=1}^N \left|{\vct s_k}\right|^2 \Biggr] }
                     \nonumber\\
&=& \int d^2{\vct X} \,
    \delta\left[(A - 1) - \left|\vct X\right|^2
                        - \sum_{k=1}^N \left|{\vct s_k}\right|^2 \right] \,
    \delta^2 \left(\vct X - \sum_{k=1}^N {\vct s_k} \right) \\
&=& {1 \over (2\pi)^3} \int d^2{\vct X} \int dv \int d^2{\vct u} \,
    \exp \biggl[-i v \bigl(w - X^2 - \sum_{k=1}^N s_k^2\bigr)
                -i {\vct u}\cdot\bigl(\vct X - \sum_{k=1}^N {\vct s}_k\bigr)
         \biggr] , \nonumber
\end{eqnarray}
where $\delta^2$ is the two-dimensional delta function and $w = A-1$.
Substituting equations (B3) and (B5) into equation (B2), we can perform
the integration with respect to $\vct s_k$:
\begin{equation}
\int d^2{\vct s}_k \, \exp [i v s_k^2 -
                            i ({\vct t}_k - {\vct u})\cdot{\vct s}_k]
= {i\pi \over v} \exp(-i |\vct t_k -\vct u|^2/4v) ,
\end{equation}
and then the integration with respect to $\vct t_k$:
\begin{equation}
{1 \over (2\pi)^2} \int d^2{\vct t}_k \, q({\vct t}_k) \,
                   \left(i\pi\over v\right)\, \exp(-i |\vct t_k -\vct u|^2/4v)
= 1 - {f(v,u) \over N} ,
\end{equation}
where
\begin{eqnarray}
f(v,u) &=& {i \tau \over 2 v} e^{-i u^2/4v} \int_0^{\infty} dt\,
           J_0\left(t u \over 2v\right) \, t^2 e^{-i t^2/4v}
\nonumber \\ & & \\
       &=& \tau \sqrt{\pi |v|\over 2} \left(1 - i {v\over |v|}\right)\,
           g(z) ,
\nonumber
\end{eqnarray}
and $J_0$ is the zero-order Bessel function.
In equation (B8), $z = i u^2/8v$,
\begin{equation}
g(z) = \left[(1+2z) I_0(z) + 2z I_1(z)\right] e^{-z},
\end{equation}
and $I_0$ and $I_1$ are the modified Bessel functions.

We can now group together the $N$ integrations over ${\vct t}_k$ (eq.[B7])
to obtain
\begin{equation}
\left[1 - {f(v,u) \over N}\right]^N = e^{-f(v,u)}
\end{equation}
in the limit $N \to \infty$.
Then
\begin{equation}
P(\delta A) = {1 \over (2\pi)^3} \int d^2{\vct X} \int dv \int d^2{\vct u} \,
              \exp \left[-i v (w - X^2)
                         -i {\vct u}\cdot{\vct X} - f(v,u)\right] .
\end{equation}
In equation (B11), the integration with respect to $\vct X$ is
\begin{equation}
\int d^2{\vct X} \, e^{i (v X^2 - {\vct u}\cdot{\vct X})}
= {i\pi \over v} e^{-iu^2/4v} ,
\end{equation}
and we are left with two integrations with respect to $v$ and $u$.
Substituting equations (B8) and (B12) into equation (B11), and using
variables $v$ and $z$ instead of $v$ and $u$,
we obtain
\begin{equation}
P(\delta A) = {2\over \pi} \int_0^\infty dz \, e^{-2z} \int_0^\infty dv \,
              \cos\left[w v - \tau \sqrt{\pi v \over 2} g(z)\right] \,
              \exp\left[-\tau \sqrt{\pi v \over 2} g(z)\right] .
\end{equation}
Finally, after performing the integration with respect to $v$, we have
$P(\delta A)$ in the form
\begin{equation}
P(\delta A) = {\tau\over w^{3/2}} \int_0^{\infty} dz\, g(z) \,
              \exp\left[-2 z - {\pi \tau^2 \over 4 w} g^2(z)\right] ,
\end{equation}
where $w=\delta A$ and $g(z)$ is defined in equation (B9).
Equation (B14) must be integrated numerically in general, but it has the
following analytic asymptotes:
\begin{equation}
\begin{array}{rcll}
P(\delta A) &=& {{\textstyle 2} \over {\textstyle \pi \tau^2}} \, y^{-1/2}
                e^{-\pi/4y} & \qquad {\rm for}~y = \delta A/\tau^2 \ll 1 ,\\
            &=& {{\textstyle 1} \over {\textstyle \sqrt{2} \tau^2}} \,
                y^{-3/2} & \qquad {\rm for}~y = \delta A/\tau^2 \gg 1 .
\end{array}
\end{equation}

\clearpage

\figcaption[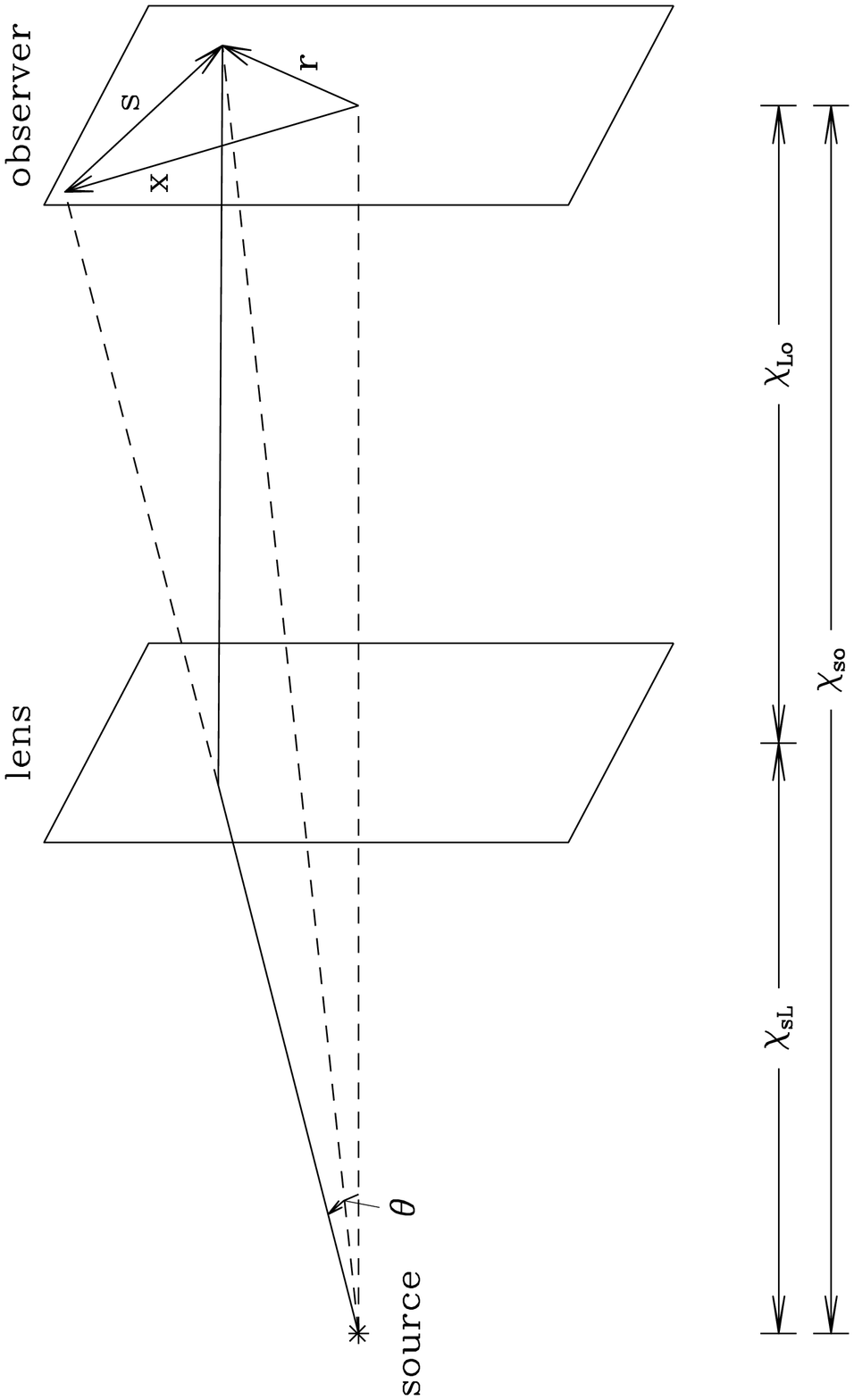]{A schematic diagram illustrating our notation for single
plane gravitational lensing.
A light ray that leaves the source with angle $\vct\theta$ has Lagrangian
coordinates $\vct x$.
It is deflected on the lens plane and reaches the observer plane
at Eulerian coordinates $\vct r$.}

\figcaption[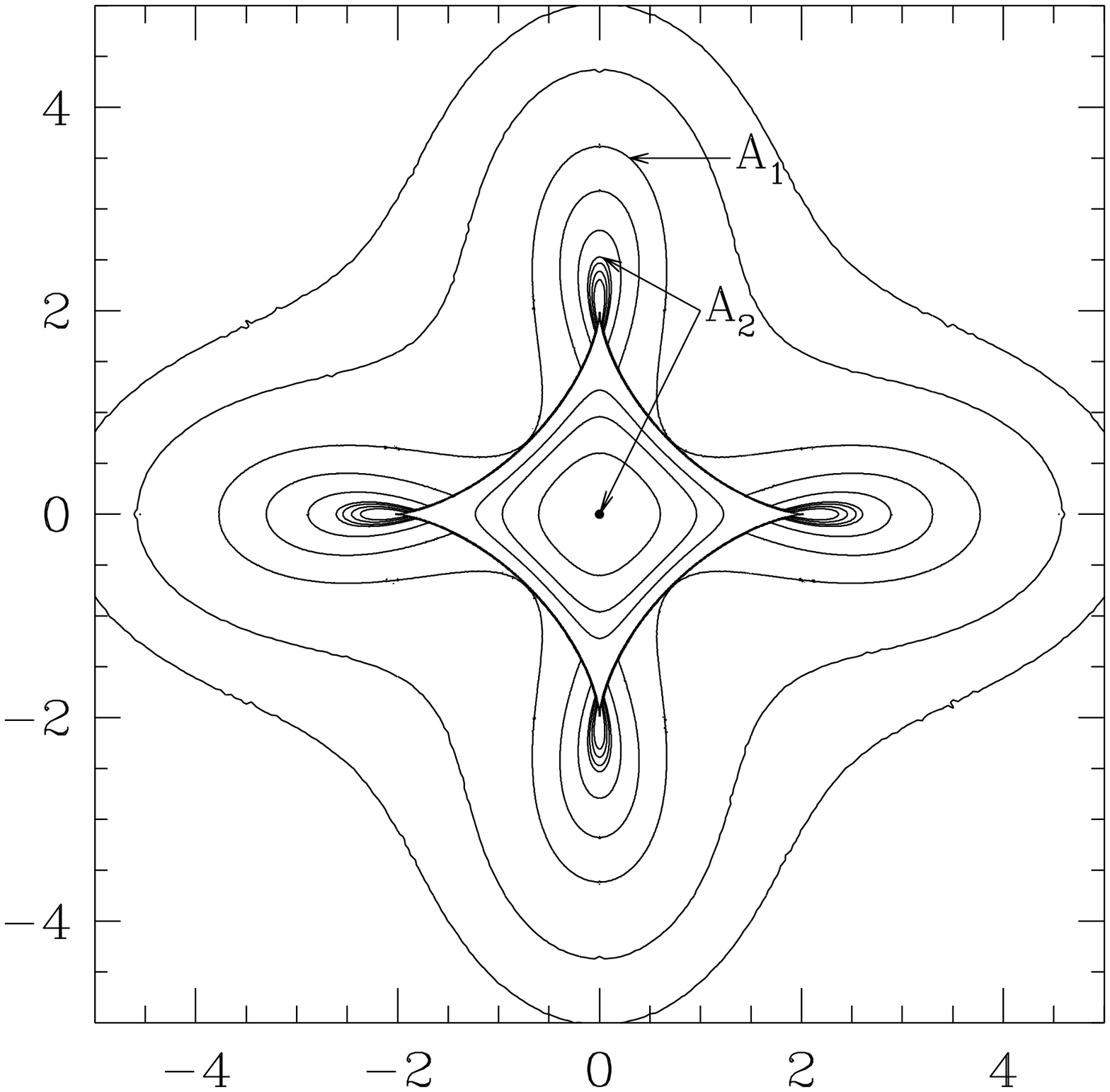]{The caustic and the iso-magnification contours on the
observer plane for the point mass plus weak shear lens.
They were obtained numerically for shear $S = 0.02$.
The axes are in units of $x_E S$. The caustic has the shape of an astroid
with the four cusps at a distance $2 x_E S$ from the origin.
The contours of critical values $A_1 = 2/3 \protect\sqrt{3} S$ and $A_2 = 1/S$
are also indicated.}

\figcaption[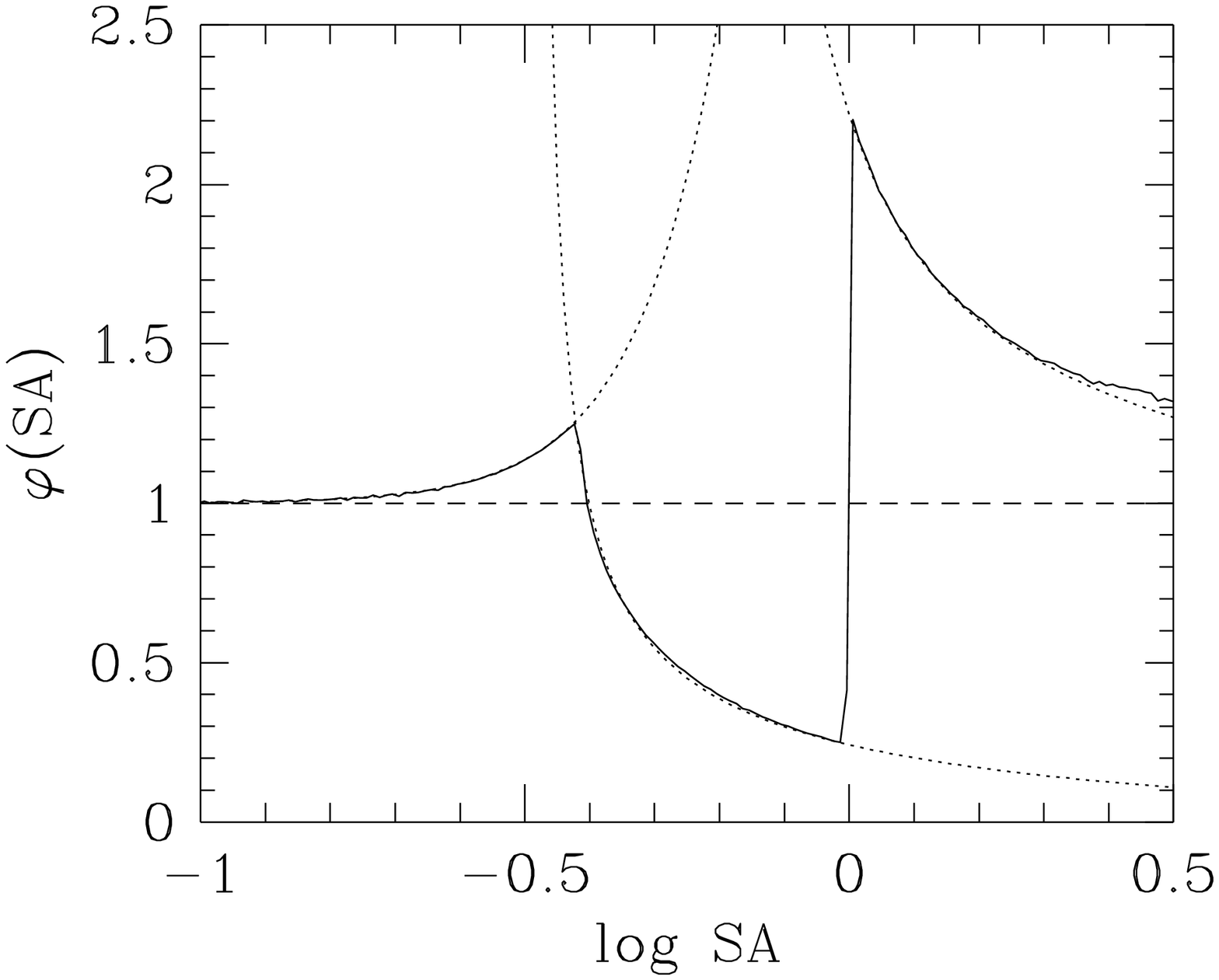]{The ``normalized'' differential cross-section,
$\varphi(S A) = \sigma(A|S)/\sigma_0(A)$, of the point mass plus weak shear
lens ({\it solid line}). It was obtained numerically for shear $S = 0.02$.
The dotted lines show the analytic fit, equation (\ref{eq28}).}

\figcaption[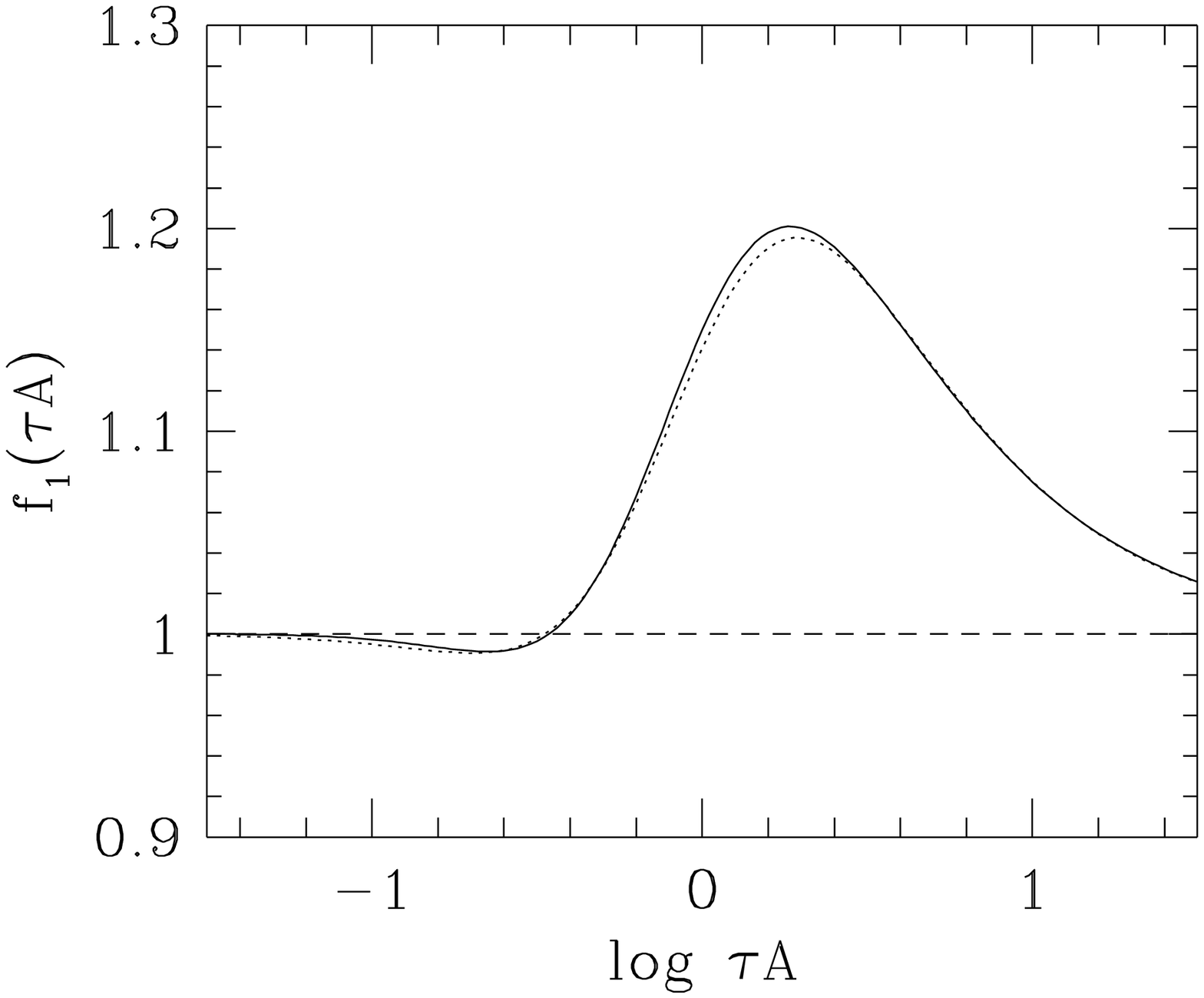]{The function $f_1 (\tau A)$ (eq.[\ref{eq32}]) that
describes the caustic-induced feature in the macroimage magnification
distribution, $P(A)$, at low optical depth $\tau$.
The solid line is the numerical result using $\varphi(S A)$ shown in
Fig.~3, and the dotted line is the analytic fit, equation (\ref{eq33}).}

\figcaption[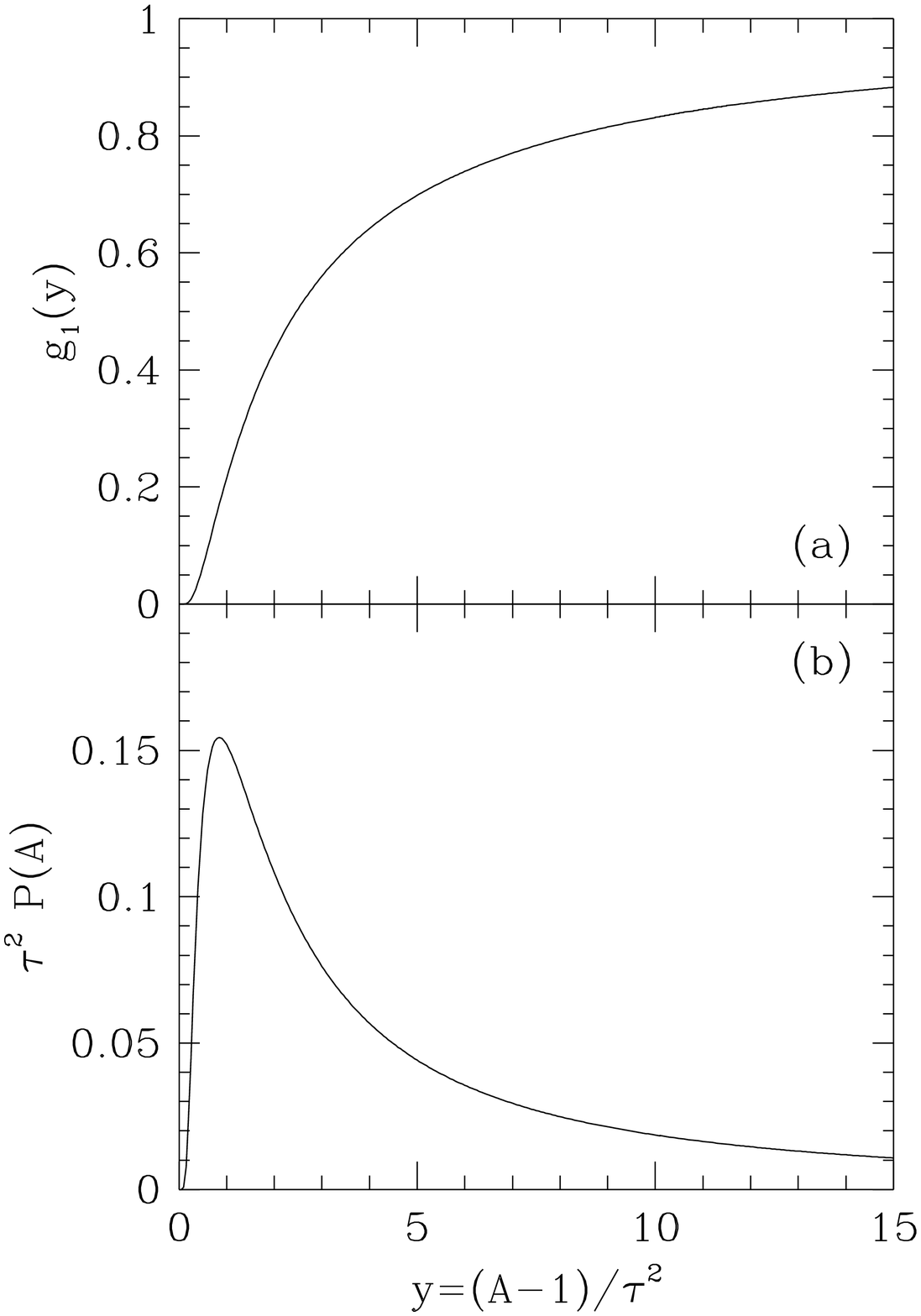]{({\it a}) The function $g_1 [y=(A-1)/\tau^2]$
(eq.[\ref{eq37}]) that describes the modification of $P(A)$ at low
magnification in the low optical depth limit.
({\it b}) The macroimage magnification distribution $P(A)$
($= g_1[y]/\protect\sqrt{2} y^{3/2} \tau^2$) at low magnification in the low
optical depth limit.}

\figcaption[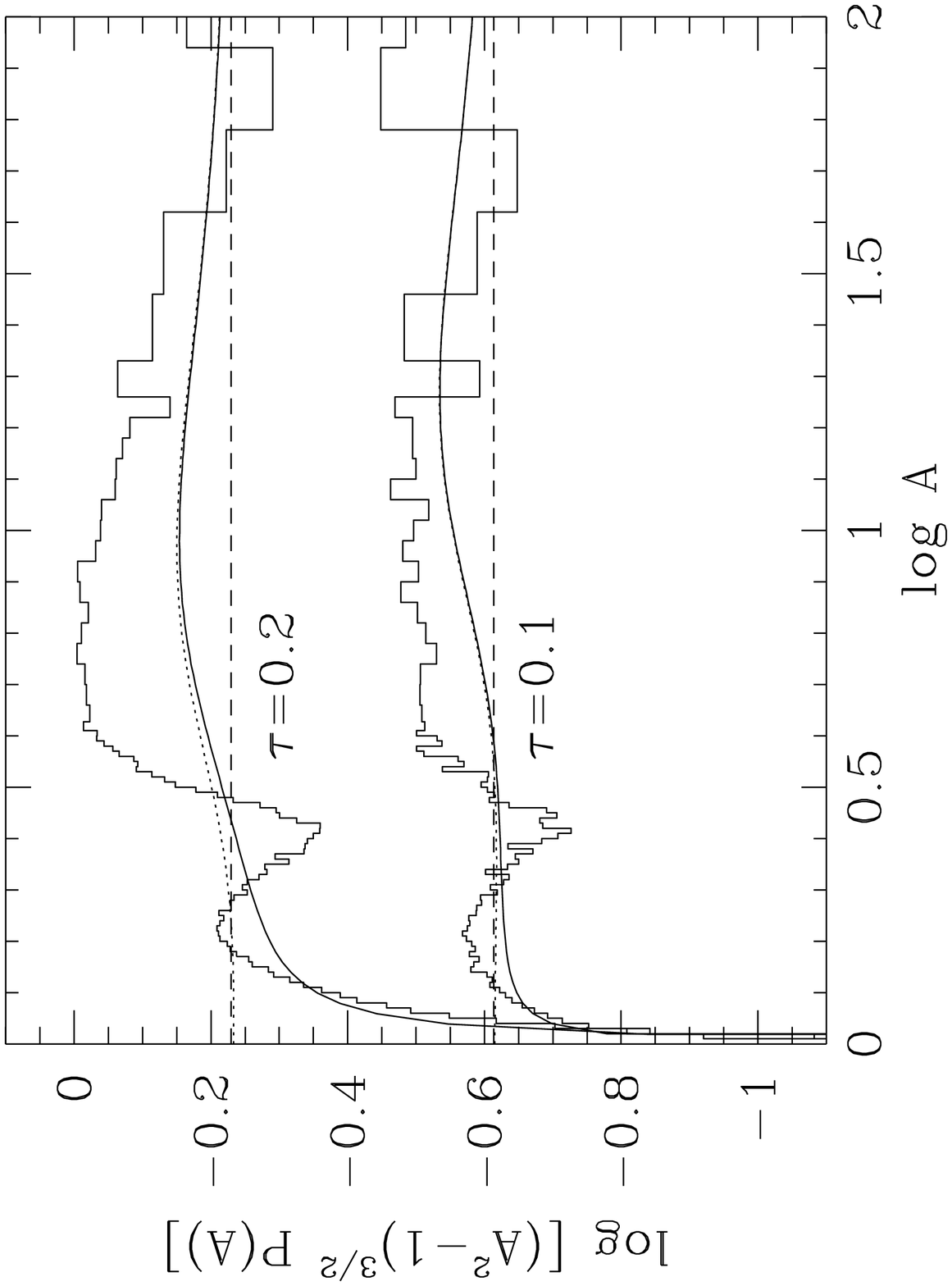]{Comparison of the semi-analytic macroimage magnification
distribution, $P(A)$, with the Monte Carlo results of Rauch \etal\ (1992)
for optical depth $\tau = 0.1$ and $0.2$ ({\it histograms}).
The solid lines are the semi-analytic $P(A)$ with the caustic-induced
feature and the modification at low $A$ (eq.[\ref{eq43}]), and the dotted
lines are the semi-analytic $P(A)$ with the caustic-induced feature only
(i.e., eq.[\ref{eq43}] without the function $g_1$).}

\clearpage

\plotone{fig1.ps}

\clearpage

\plotone{fig2.ps}

\clearpage

\plotone{fig3.ps}

\clearpage

\plotone{fig4.ps}

\clearpage

\plotone{fig5.ps}

\clearpage

\plotone{fig6.ps}

\end{document}